\documentclass[10pt,conference]{IEEEtran}
\IEEEoverridecommandlockouts
\usepackage{cite}
\usepackage{amsmath,amssymb,amsfonts}
\usepackage{algorithmic}
\usepackage{graphicx}
\usepackage{hyperref}
\usepackage{booktabs}
\usepackage{textcomp}
\usepackage{xcolor}
\usepackage{multirow}
\def\BibTeX{{\rm B\kern-.05em{\sc i\kern-.025em b}\kern-.08em
    T\kern-.1667em\lower.7ex\hbox{E}\kern-.125emX}}
\begin{document}

\title{Do Code LLMs Understand Design Patterns?}

\author{
\IEEEauthorblockN{Zhenyu Pan}
\IEEEauthorblockA{
\textit{Northwestern University}\\
zhenyupan@u.northwestern.edu}
\and
\IEEEauthorblockN{Xuefeng Song}
\IEEEauthorblockA{
\textit{Northwestern University}\\
\quad xuefengsong2026@u.northwestern.edu \quad}
\and
\IEEEauthorblockN{Yunkun Wang}
\IEEEauthorblockA{
\textit{Zhejiang University}\\
\quad wangykun@zju.edu.cn \quad}

\and
\IEEEauthorblockN{Rongyu Cao}
\IEEEauthorblockA{
\textit{Alibaba Group}\\
caorongyu.cry@alibaba-inc.com}
\and
\IEEEauthorblockN{Binhua Li}
\IEEEauthorblockA{
\textit{Alibaba Group}\\
binhua.lbh@alibaba-inc.com}
\and
\IEEEauthorblockN{Yongbin Li}
\IEEEauthorblockA{
\textit{Alibaba Group}\\
shuide.lyb@alibaba-inc.com}
\and
\IEEEauthorblockN{Han Liu}
\IEEEauthorblockA{
\textit{Northwestern University}\\
hanliu@northwestern.edu}
}
\maketitle

\begin{abstract}
Code Large Language Models (LLMs) demonstrate great versatility in adapting to various downstream tasks, including code generation and completion, as well as bug detection and fixing. However, Code LLMs often fail to capture existing coding standards, leading to the generation of code that conflicts with the required design patterns for a given project. As a result, developers must post-process to adapt the generated code to the project's design norms. In this work, we empirically investigate the biases of Code LLMs in software development. Through carefully designed experiments, we assess the models' understanding of design patterns across recognition, comprehension, and generation. Our findings reveal that biases in Code LLMs significantly affect the reliability of downstream tasks.
\end{abstract}

\section{Introduction}
LLMs transform traditional paradigms across various fields \cite{b20}. Among these \cite{b21}, Code LLMs, trained on extensive code data, demonstrate great versatility in code-related tasks. They facilitate automated software development through downstream tasks such as code generation and completion, bug localization and repair, and security detection and enhancement, which significantly improves developer productivity.

However, there are still considerable challenges in applying code LLMs in real-world scenarios, especially regarding specific aspects of software engineering. For instance, in complex repository environments, it is difficult to extract the necessary background knowledge from a large amount of information and follow project-specific conventions during code generation. Current code LLMs often fail to properly understand the existing design patterns and coding styles of a project, leading to generated code that does not meet project requirements. This mismatch creates an additional burden for developers, such as requiring extensive code revisions and post-processing.

Recent empirical studies primarily evaluate code LLMs on tasks like code generation, bug detection, and code summarization, often using benchmarks such as CodeXGLUE \cite{b11,b12,b13}. While these analyses have provided insights into generation accuracy and contextual understanding, the role of design patterns in software engineering has been largely overlooked. Design patterns are essential for maintaining software quality, particularly in object-oriented development. Emphasizing bias analysis is crucial, especially regarding potential biases in code LLMs under different design patterns, as such biases may impact the reliability of downstream tasks.

To bridge this gap, we propose an empirical study to evaluate code LLMs' understanding and generation of design patterns, focusing on whether these models can classify and generate code while adhering to established patterns like Singleton and Factory. For our dataset, we manually selected high-quality repositories from GitHub for both Python and Java, each corresponding to 12 design patterns, with two repositories chosen for each pattern in each language. Based on this, we designed three experiments to assess the capabilities of Code LLMs in recognition, generation, and comprehension of design patterns. For recognition, we conducted a Design Pattern Classification experiment to classify the design patterns in a given code file. For generation and comprehension, we designed two sets of experiments involving code completion and function generation, both with and without prior information about the design pattern. Based on the evaluation of these three capabilities, we provide an analysis of the strengths and weaknesses of Code LLMs in handling design patterns.

The main contributions of our work are : (1) We propose an evaluation framework for recognizing, generating, and understanding design patterns, (2) Through experiments, we demonstrate that code LLMs exhibit biases when adhering to specific design patterns. (3) We illustrate how these biases affect the reliability of code generation and the additional workload imposed on developers. (4) We emphasize the significance of our work for future research and practical applications, such as improving model training and guiding models to better follow development standards.

\section{Related Work}

\subsection{Code LLMs}
Code LLMs demonstrate notable advancements, each with distinct strengths and limitations. The GPT-4 series excel in reasoning, multi-modal tasks, and code generation, but requires high computational resources and shows variability across updates \cite{b1}. Claude 3.5 focuses on safety and ethics in code generation, performing well in simple tasks but struggling with complex logic \cite{b3}. CodeQwen is efficient in domain-specific tasks like structured retrieval and technical analysis but lacks adaptability for general use \cite{b4}. The LLaMA 3.1 series are open-source, emphasizing efficiency and accessibility, but their reliance on public datasets limits generalization in niche scenarios \cite{b5}. StarCoder \cite{b17} excels in code generation but needs significant resources and has limited support for niche languages. Mistral is optimized for lightweight tasks and excels in long-context scenarios with its 128k context window, though it lacks multi-modal capabilities \cite{b7}. Yi is strong in algorithmic tasks and technical problem-solving but struggles with multitasking and long-context efficiency \cite{b8}. The Qwen series is scalable and efficient in structured applications but less effective for general-purpose tasks \cite{b9}.

\subsection{Empirical Studies on Code LLMs}
Recent empirical studies on code LLMs focus on evaluating code generation, bug detection, code summarization, and automated documentation \cite{b19}. Researchers explore factors such as model architecture variations, the diversity and size of training datasets, and fine-tuning techniques to understand their impact on the performance of code LLMs in handling diverse code-related tasks \cite{b15}. Additionally, they investigate the models' ability to comprehend contextual information, maintain code consistency, and manage complex programming constructs \cite{b16}. Despite these comprehensive analyses, the role of design patterns in software engineering remains underexplored. Design patterns are crucial in software development, particularly in object-oriented programming, as they enhance the overall quality and maintainability of projects. However, there has been limited analysis of code LLMs' abilities to work with design patterns. To address this gap, we propose a focused study on code LLMs' understanding of design patterns. Specifically, we aim to assess whether code LLMs can accurately classify code files based on established design patterns, such as Singleton and Factory, and evaluate their ability to complete code while preserving these patterns. By measuring the similarity between generated and original code, our research seeks to provide deeper insights into the capabilities of code LLMs in applying software design patterns.
\vspace{-0.15in}
\section{Design Pattern Evaluation}
\vspace{-0.05in}
\subsection{Experiment design}
\subsubsection{Overview}
We evaluate the design pattern comprehension ability of code LLMs in different tasks across two programming languages, Python and Java: (i) \textbf{Design Pattern Classification}, which assesses the model’s ability to identify and classify different object-oriented design patterns, such as Singleton, Factory, and Observer, within given code snippets. The objective is to determine if the model can accurately recognize structural and behavioral characteristics that define each design pattern in both Python and Java; (ii) \textbf{Line Completion}, where the model completes missing lines within a code snippet that follows a particular design pattern. This task tests the model’s understanding of the context and requirements of specific design patterns, as it must generate lines that adhere to the pattern’s conventions in each language; and (iii) \textbf{Function Generation}, which evaluates the model’s ability to generate entire functions with given description in with/without given design pattern, assessing whether the model can recognize and apply structural and procedural elements of the pattern to fulfill functional requirements effectively in both Python and Java.

\subsubsection{Models}
Our study draws upon a wide range of state-of-the-art large language models (LLMs), each selected to showcase distinct strengths across various model families and architectural designs. We employ general-purpose models, such as GPT-4 and Claude, alongside code-focused models like GLM and DeepSeek, which are tailored for tasks involving code understanding and generation. Additionally, we include models from specific series, such as Qwen and LLaMA, and incorporate multimodal and multilingual capabilities through options like Yi and Mistral. This diverse model set enables a robust evaluation of LLM performance on code-centric tasks, with a particular focus on assessing capabilities in design pattern comprehension within domain-specialized applications.

\subsubsection{Datasets}
We manually selected 48 high-quality repositories from GitHub, with 24 in Python and 24 in Java, representing 12 design patterns. Each pattern was represented by two repositories per language, and the code files were categorized into three levels based on complexity: simple, moderate, and difficult. For the Design Pattern Classification task, we provided code files to LLMs to determine the design pattern. In the Line Completion task, we randomly removed three separate lines of effective code and asked the LLMs to complete it based on the context. For the Function Generation task, we extracted a function, used GPT-4 to write a description, and provided the context and description to the LLMs to generate the function. We test both with and without given design pattern to evaluate the models' understanding.
\subsubsection{Metrics}
For the classification results, we used simple accuracy to evaluate the models' performance.
For the code completion and generation tasks, we used two metrics: code similarity (CS) and code edit similarity (ES).
CS measures the overlap between the generated code and the reference code. We use difflib's SequenceMatcher, which employs the Ratcliff/Obershelp algorithm to finds the Longest Matching Subsequence (LCS) between two sequences and then recursively processes the unmatched parts to calculate the similarity ratio. ES measures the minimal number of edit operations (insertion, deletion, substitution) required to transform the generated code into the reference code, normalized by the length of the reference.

\subsection{Evaluation on Design Pattern Classification}

\subsubsection{Empirical Analysis}
\begin{table*}[h]
\centering
\caption{Results of design pattern classification by different LLMs on Java and Python code files of varying complexity}
\vspace{-0.1in}
\label{tab:all-acc}
\resizebox{\textwidth}{!}{
\begin{tabular}{l|c|c|c|c|c|c|c|c|c|c|c}
\toprule
\multirow{2}{*}{\textbf{Dataset}} & \multicolumn{2}{c|}{\textbf{OpenAI Models}} & \multicolumn{2}{c|}{\textbf{Qwen Models}} & \multicolumn{2}{c|}{\textbf{Llama Models}} & \multicolumn{5}{c}{\textbf{Other Models}} \\
\cline{2-12}
& \textbf{GPT-4} & \textbf{GPT-4o} & \textbf{CodeQwen} & \textbf{Qwen\_25\_72B} & \textbf{Llama\_31\_405B} & \textbf{Llama\_31\_70B} & \textbf{DeepSeek\_V2} & \textbf{Claude35} & \textbf{Mistral\_123B} & \textbf{Yi\_34B} & \textbf{GLM\_3\_6B} \\
\hline
Java Easy      & 64.29 & 71.43 & 42.86 & 71.43 & 64.29 & 71.43 & 64.29 & 42.86 & 57.14 & 57.14 & 57.14 \\
\hline
Java Medium    & None & 14.29 & None & 28.57 & 28.57 & 14.29 & None & 14.29 & None & 14.29 & 14.29 \\
\hline
Java Hard      & 33.33 & 44.44 & 22.22 & 22.22 & 33.33 & 55.56 & 33.33 & 44.44 & 44.44 & 33.33 & 11.11 \\
\hline
Java All       & 40 & 50 & 26.67 & 46.67 & 46.67 & 53.33 & 40 & 36.67 & 40 & 40 & 33.33 \\
\hline
Python Easy    & 33.33 & 44.44 & 44.44 & 33.33 & 33.33 & 33.33 & 22.22 & 11.11 & 44.44 & 44.44 & 44.44 \\
\hline
Python Medium  & 35.71 & 42.86 & 42.86 & 35.71 & 35.71 & 28.57 & 35.71 & 35.71 & 35.71 & 28.57 & 21.43 \\
\hline
Python Hard    & None & 7.14 & None & 7.14 & 14.29 & 21.43 & 7.14 & 14.29 & 7.14 & 14.29 & 14.29 \\
\hline
Python All    & 21.62 & 29.73 & 27.03 & 24.32 & 27.03 & 27.03 & 21.62 & 21.62 & 27.03 & 27.03 & 24.32 \\
\hline
All            & 29.85 & 38.81 & 26.87 & 34.33 & 35.82 & 38.81 & 29.85 & 28.36 & 32.84 & 32.84 & 28.36 \\
\bottomrule
\end{tabular}
\vspace{-0.6in}
}
\end{table*}

\begin{figure*}
    \centering
    \vspace{-0.15in}
    \includegraphics[width=1\linewidth]{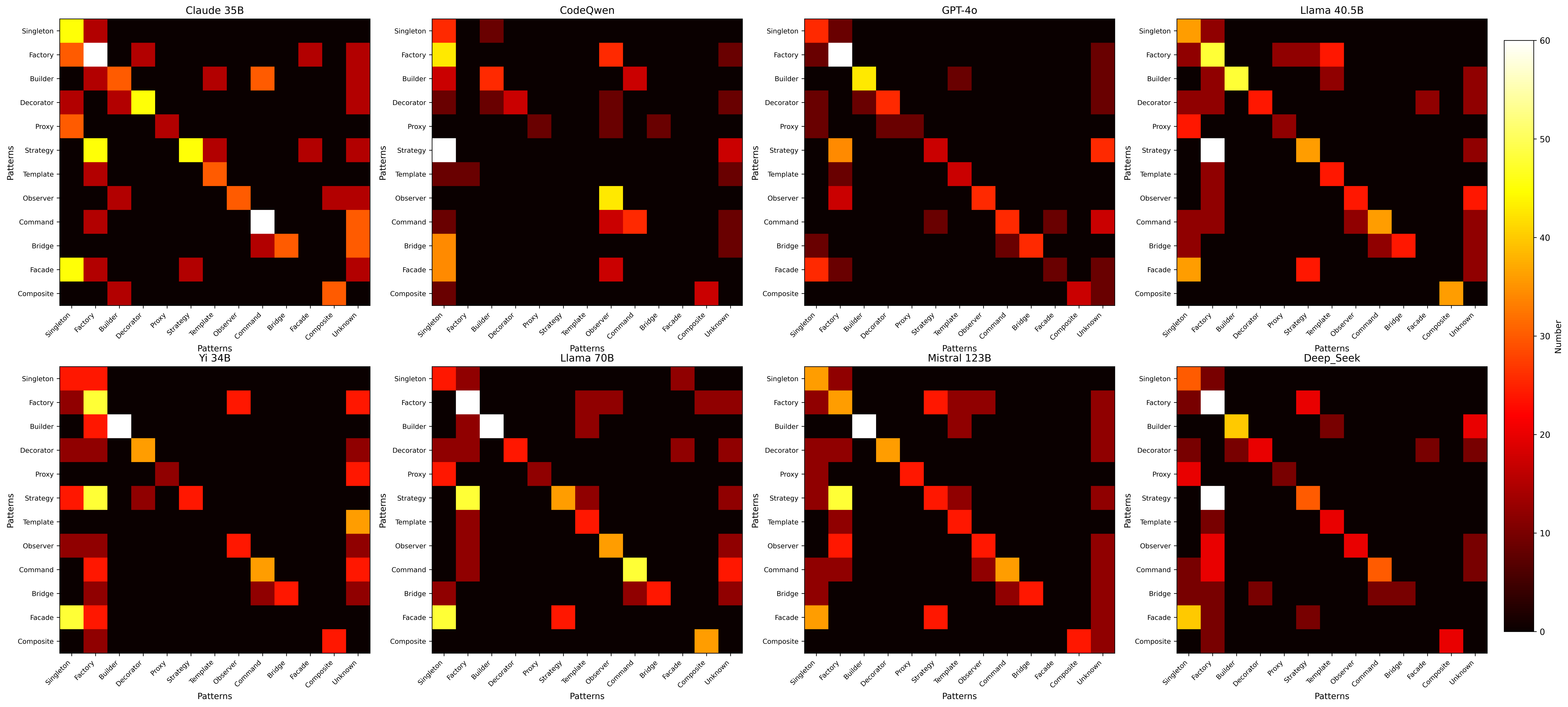}
    \vspace{-0.3in}
    \caption{Heatmaps of misclassifications by 8 LLMs across 12 design patterns involved in the experiments. The X-axis includes an additional category representing cases where the LLMs failed to classify the design pattern.}
    \label{fig:heatmap}
    \vspace{-0.2in}
\end{figure*}

As shown in Table~\ref{tab:all-acc}, GPT-4o and Llama-31-70B achieved the highest overall accuracy, both scoring 38.81\% across all datasets. In the Java datasets, Llama-31-70B performed best overall with 53.33\% accuracy on Java All, and it excelled in the Java Hard subset with 55.56\%. Both GPT-4o and Llama-31-70B achieved the highest accuracy on Java Easy at 71.43\%. For the Python datasets, GPT-4o led with an overall accuracy of 29.73\% on Python All. On Python Easy, several models, including CodeQwen, Mistral-123B, Yi-34B, and GLM-3-6B, matched high accuracy levels at 44.44\%. A general trend across all models is a decrease in performance from Java to Python and from Easy to Hard datasets. This suggests that the LLMs find Python code and more complex code samples more challenging, possibly due to differences in language syntax or less exposure to complex patterns during training. In summary, while GPT-4o and Llama-31-70B stand out in overall performance, all models show room for improvement, especially in handling complex Python code.

From the heat map shown in Figure~\ref{fig:heatmap}, when using LLMs to identify design patterns in code, Singleton and Factory patterns are often over-predicted or misattributed due to their prevalence and recognizable structure. Singleton involves centralized object management, which overlaps with patterns like Facade, Proxy, or Command, leading to frequent misclassification. Factory, similarly, is confused with Abstract Factory, Builder, or Strategy due to shared object creation logic. Frameworks like Spring (Java) or Python’s dependency injection further blur these distinctions, making it hard for LLMs to discern intent. Real-world implementations often combine patterns, adding to the challenge. LLMs heuristically rely on static methods, single-instance management, or object creation cues, which can align with Singleton or Factory but are not exclusive to them. Ambiguity in natural language prompts and insufficient training data also contribute to frequent misclassifications. Addressing these issues requires improved training data, clearer distinctions, and refined contextual understanding of overlapping features in code. LLMs also struggle with the Facade pattern due to its abstract nature and contextual dependency. Unlike Singleton or Factory, which have distinct structural markers, Facade simplifies access to complex subsystems via a unified interface, making its identification challenging. Its reliance on naming conventions and subtle structural cues, along with integration with other dominant patterns, often masks its intent. This complexity, combined with insufficient training examples, makes Facade one of the harder patterns for LLMs to reliably identify.
\begin{table*}
\centering
\caption{Overall comparison of different LLMs on line completion and function generation with and without prior knowledge of design pattern. Each result box shows line completion on the left and function generation on the right, separated by a '/'.}
\vspace{-0.1in}
\label{tab:line_overall}
\resizebox{\textwidth}{!}{
\begin{tabular}{|c|c|c|c|c|c|c|c|c|c|c|c|c|c|c|c|c|c|c|c|c|c|c|}
\toprule
\multirow{2}{*}{\textbf{Models}} & \multicolumn{4}{c|}{\textbf{OpenAI Models}} & \multicolumn{4}{c|}{\textbf{Qwen Models}} & \multicolumn{4}{c|}{\textbf{Llama Models}} & \multicolumn{10}{c|}{\textbf{Other Models}} \\ 
\textbf{} & \multicolumn{2}{c|}{\textbf{GPT-4}} & \multicolumn{2}{c|}{\textbf{GPT-4o}} & \multicolumn{2}{c|}{\textbf{CodeQwen}} & \multicolumn{2}{c|}{\textbf{Qwen\_25\_72B}} & \multicolumn{2}{c|}{\textbf{Llama\_31\_405B}} & \multicolumn{2}{c|}{\textbf{Llama\_31\_70B}} & \multicolumn{2}{c|}{\textbf{DeepSeek\_V2}} & \multicolumn{2}{c|}{\textbf{Claude35}} & \multicolumn{2}{c|}{\textbf{Mistral\_123B}} & \multicolumn{2}{c|}{\textbf{Yi\_34B}} & \multicolumn{2}{c|}{\textbf{GLM\_3\_6B}} \\ 
\hline
\textbf{Design Pattern Provided (Y/N)} & Yes & No & Yes & No& Yes & No& Yes & No& Yes & No& Yes & No& Yes & No& Yes & No& Yes & No& Yes & No& Yes & No \\ 
\hline
\textbf{Code Similarity (\%)}  & 24.68/23.23 & 22.99/31.05 & 37.35/32.93 & 51.31/53.71 & 5.4/11.20 & 5.50/10.79 & 20.27/39.67 & 17.84/42.75 & 37.02/28.08 & 41.02/44.70 & 26.59/29.67 & 26.49/41.55 & 18.01/19.89 & 12.11/30.90 & 45.75/32.45 & 48.16/33.63 & 29.83/10.02 & 40.16/39.58 & 8.17/14.65 & 9.30/18.20 & 7.99/10.65 & 8.61/10.21 \\ 
\hline
\textbf{Edit Similarity (\%)} & 23.91/26.08 & 21.21/32.08 & 33.30/32.77 & 43.72/54.27 & 6.2/12.21 & 5.54/13.74 & 20/39.11 & 17.72/42.34 & 32.58/25.75 & 34.78/45.73 & 24.36/28.97 & 24.98/43.21 & 18.48/21.01 & 13.54/31.41 & 37.35/29.91 & 41.50/33.40 & 28.31/16.51 & 35.22/38.62 & 8.90/19.17 & 9.96/23.99 & 11.39/15.60 & 12.60/16.06 \\ 
\bottomrule
\end{tabular}
}
\vspace{-0.2in}
\end{table*}

\begin{table*}
\centering
\caption{Comparison of LLMs on function generation for various design patterns with and without prior knowledge of the code file’s design pattern. Each result box shows code similarity (\%) on the left and edit similarity (\%) on the right, separated by a '/'.}
\vspace{-0.1in}
\label{tab:pattern}
\resizebox{\textwidth}{!}{
\begin{tabular}{|c|c|c|c|c|c|c|c|c|c|c|c|c|c|c|c|c|c|c|c|c|c|c|}
\toprule
\multirow{2}{*}{\textbf{Models}} & \multicolumn{4}{c|}{\textbf{OpenAI Models}} & \multicolumn{4}{c|}{\textbf{Qwen Models}} & \multicolumn{4}{c|}{\textbf{Llama Models}} & \multicolumn{10}{c|}{\textbf{Other Models}} \\ 
\textbf{} & \multicolumn{2}{c|}{\textbf{GPT-4}} & \multicolumn{2}{c|}{\textbf{GPT-4o}} & \multicolumn{2}{c|}{\textbf{CodeQwen}} & \multicolumn{2}{c|}{\textbf{Qwen\_25\_72B}} & \multicolumn{2}{c|}{\textbf{Llama\_31\_405B}} & \multicolumn{2}{c|}{\textbf{Llama\_31\_70B}} & \multicolumn{2}{c|}{\textbf{DeepSeek\_V2}} & \multicolumn{2}{c|}{\textbf{Claude35}} & \multicolumn{2}{c|}{\textbf{Mistral\_123B}} & \multicolumn{2}{c|}{\textbf{Yi\_34B}} & \multicolumn{2}{c|}{\textbf{GLM\_3\_6B}} \\ 
\hline
\textbf{Design Pattern Provided (Y/N)} & Yes & No & Yes & No& Yes & No& Yes & No& Yes & No& Yes & No& Yes & No& Yes & No& Yes & No& Yes & No& Yes & No \\ 
\hline
\textbf{Singleton}  & 33.64/39.89 & 13.20/32.10 & 36.15/32.77 & 42.73/50.79 & 2.74/5.47 & 2.48/4.98 & 52.57/44.11 & 35.19/36.03 & 35.08/33.22 & 32.52/58.02 & 37.37/34.83 &39.60/49.90 & 18.55/14.65 & 19.77/19.78 & 22.82/21.85 & 12.91/33.17 & 6.63/11.40 & 62.12/55.73 & 23.50/21.19 & 13.78/22.92 & 19.07/22.12 & 7.94/15.22 \\ 
\hline
\textbf{Factory} & 34.31/34.23 & 34.72/30.64 & 20.22/23.34 & 47.92/52.73 & 14.57/15.52 & 8.99/11.65 & 25.82/31.21 & 31.50/29.85 & 32.55/32.66 & 58.14/54.32 & 29.24/28.42 & 36.80/34.45 & 14.76/20.60 & 23.06/25.77 & 32.47/33.00 & 15.53/19.52 & 7.72/17.70 & 25.30/26.07 & 6.55/15.98 & 8.92/12.92 & 6.55/15.16 & 2.69/8.58 \\ 
\hline
\textbf{Builder} & 6.68/21.07 & 13.41/26.04 & 28.12/41.23 & 33.89/43.64 & 14.41/16.52 & 15.13/20.81 & 10.87/21.00 & 9.31/17.46 & 15.90/24.81 & 34.07/40.58 & 6.39/19.64 &13.94/22.96 & 3.22/11.00 & 9.47/15.80 & 11.70/24.41 & 17.47/27.47 & 8.57/17.39 & 12.77/17.98 & 9.76/25.77 & 10.91/30.16 & 5.13/15.32 & 4.95/13.52 \\ 
\hline
\textbf{Decorator} & 31.44/35.37 & 41.15/31.48 & 33.10/32.20 & 62.03/62.57 & 17.75/24.17 & 13.24/23.33 & 41.55/36.47 & 61.06/61.04 & 32.67/24.83 & 52.00/49.47 & 32.01/32.19 &53.69/52.46 & 30.04/36.05 & 52.81/52.24 & 36.84/33.21 & 49.56/53.39 & 13.81/20.76 & 37.55/35.70 & 19.45/26.41 & 25.59/33.44 & 24.10/23.82 & 30.20/27.96 \\ 
\hline
\textbf{Proxy} & 35.48/34.06 & 27.18/20.18 & 36.90/37.00 & 36.22/32.08 & 27.69/28.86 & 9.77/14.50 & 36.48/35.62 & 36.70/32.98 & 29.37/27.22 & 35.16/30.65 & 28.38/24.52 &35.08/30.62 & 21.08/15.53 & 33.44/31.18 & 26.94/23.56 & 26.23/19.31 & 5.72/14.46 & 34.03/29.30 & 4.71/15.56 & 6.01/4.95 & 12.99/15.49 & 2.82/12.13 \\ 
\hline
\textbf{Strategy} & 29.60/28.45 & 43.32/43.18 & 37.28/35.75 & 65.08/60.22 & 7.39/8.44 & 13.76/11.65 & 37.91/37.32 & 48.61/47.48 & 25.03/23.83 & 44.20/44.36 & 20.86/21.84 &44.03/46.95 & 16.98/20.03 & 34.95/34.12 & 35.59/31.12 & 41.57/38.08 & 10.28/17.36 & 54.15/50.15 & 19.65/19.64 & 19.22/21.19 & 10.26/15.66 & 18.51/23.00 \\ 
\hline
\textbf{Template} & 33.62/33.33 & 54.19/46.37 & 62.20/51.98 & 65.59/60.11 & 10.49/7.06 & 27.37/21.76 & 66.17/61.46 & 62.40/57.79 & 49.88/37.91 & 57.13/50.01 & 44.44/36.74 & 66.45/66.09 & 30.41/26.21 & 36.75/35.58 & 59.31/43.98 & 48.59/38.56 & 16.69/18.04 & 63.12/50.14 & 2.51/6.36 & 3.09/7.10 & 8.50/14.86 & 1.95/11.44 \\ 
\hline
\textbf{Observer} & 7.67/5.86 & 5.08/5.30 & 15.46/10.29 & 76.48/70.35 & 7.96/6.58 & 6.73/4.92 & 48.23/44.82 & 77.50/64.55 & 13.79/10.30 & 70.38/62.31 & 28.48/20.16 & 43.07/40.17 & 16.23/11.56 & 43.73/29.23 & 45.53/29.67 & 36.48/23.06 & 12.21/9.09 & 48.02/45.74 & 42.08/36.55 & 41.13/34.52 & 14.19/9.68 & 9.02/7.12 \\ 
\hline
\textbf{Command} & 16.30/16.03 & 31.19/36.99 & 24.80/19.02 & 51.71/52.11 & 7.55/6.00 & 1.65/9.98 & 64.51/55.23 & 39.66/46.10 & 24.07/14.96 & 35.94/45.65 & 49.09/37.33 &36.70/42.28 & 24.88/18.05 & 37.82/45.08 & 27.24/18.61 & 49.60/40.90 & 11.08/13.43 & 37.39/44.05 & 12.67/10.79 & 32.66/38.14 & 3.68/12.40 & 8.59/15.18 \\ 
\hline
\textbf{Bridge} & 6.78/17.21 & 6.67/15.98 & 37.07/40.71 & 37.58/40.12 & 11.55/11.01 & 12.05/12.66 & 37.14/39.17 & 30.15/28.63 & 28.94/27.63 & 20.12/20.08 & 31.08/36.29 &21.33/18.23 & 32.71/30.17 & 25.15/24.73 & 28.14/28.91 & 26.54/22.94 & 3.02/16.23 & 32.19/32.84 & 8.46/21.26 & 10.85/18.09 & 5.15/7.99 & 4.03/7.24 \\ 
\hline
\textbf{Facade} & 11.29/11.67 & 39.81/41.41 & 32.54/31.91 & 56.92/56.33 & 5.22/5.12 & 3.71/5.62 & 29.16/31.38 & 39.27/39.57 & 30.78/25.05 & 53.04/50.49 & 37.99/38.90 & 59.49/61.47 & 15.44/21.76 & 19.06/23.43 & 36.57/34.69 & 41.95/37.29 & 7.76/13.57 & 29.14/32.45 & 15.13/14.72 & 18.29/24.81 & 8.01/12.65 & 3.98/15.25 \\ 
\hline
\textbf{Composite} & 11.99/18.30 & 24.57/25.52 & 35.32/32.79 & 63.52/64.07 & 4.68/3.68 & 9.23/16.08 & 58.40/56.79 & 58.96/53.79 & 19.76/23.66 & 39.68/40.56 & 32.29/25.48 & 50.59/51.61 & 24.81/18.08 & 32.78/27.15 & 34.08/32.76 & 27.06/28.30 & 16.85/20.56 & 51.22/52.05 & 17.55/16.55 & 35.39/41.27 & 9.64/14.34 & 4.59/17.10 \\ 
\bottomrule
\end{tabular}
}
\vspace{-0.25in}
\end{table*}
\textit{Insights:
The classification results indicate that even the best-performing models, GPT-4o and Llama-31-70B, achieved only 38.81\% overall accuracy, highlighting substantial room for improvement in design pattern recognition. A consistent decline in performance from Java to Python and from Easy to Hard datasets suggests that LLMs find Python code and complex samples more challenging, due to syntax differences and less exposure during training.
The heatmaps reveal that Singleton and Factory patterns are often over-predicted or misclassified. Their prevalence and distinctive structures make them default predictions, but overlaps with patterns like Facade or Strategy lead to confusion. LLMs particularly struggle with the Facade pattern due to its abstract nature and lack of explicit structural markers, making it difficult to identify without contextual understanding. These findings underscore the need for enhanced training data, clearer distinctions between patterns, and improved contextual comprehension within LLMs to increase reliability in software development tasks.}

\subsection{Evaluation on Line Completion and Function Generation}
\subsubsection{Empirical Analysis}
As shown in Table~\ref{tab:line_overall}, we find: (1) CS and ES Relationship: Higher CS correlates with higher ES, meaning that code closely matching the original typically requires less effort to adapt. However, this relationship isn't always linear. For instance, GPT-4o achieves high CS without design pattern knowledge, but its ES scores suggest significant edits may still be required, especially in line completion tasks. (2) Impact of Design Pattern Knowledge and Model Variability: Several models, including GPT-4o and Llama-31-405B, perform better without design pattern information, implying reliance on internalized patterns from training. Claude35 consistently excels in generating accurate, modifiable code, while models like CodeQwen show lower performance. Function generation often yields higher similarity scores than line completion, suggesting better model handling of larger code segments. Overall, improving design pattern integration, practical utility, and task-specific optimization could enhance LLM reliability and reduce manual editing efforts.

Table~\ref{tab:pattern} highlights that providing design pattern improves model performance, as seen with patterns like Singleton, where GPT-4 achieved higher Code Similarity (33.64\%) with pattern knowledge than without (13.20\%). However, exceptions exist; some models, such as GPT-4o, performed better without design pattern knowledge in patterns like Observer, suggesting effective reliance on internalized pattern knowledge. Performance varied across patterns, with Template and Decorator yielding higher scores, while Builder, Bridge, and Facade proved more challenging. A nuanced relationship between CS and ES was observed. 
Higher CS often corresponds with higher ES, suggesting that code closer to the original requires less effort to edit. However, there are cases where this relationship is not strong. For example, in the Decorator pattern, GPT-4 with design pattern knowledge had CS of 31.44\% and ES of 35.37\%, but without design pattern knowledge, CS increased and ES decreased. Model performance variability was evident, with GPT-4o and Llama-31-70B excelling in multiple patterns, while models like CodeQwen struggled.

\textit{Insights: Results reveal a nuanced relationship between CS and ES, indicating that high CS does not always equate to reduced editing effort for programmers. While some LLMs generate code resembling the original, these outputs may still require significant modifications due to underlying issues. It highlights the need for future Code LLM to prioritize better integration of design pattern knowledge and the generation of code that minimizes manual edits. Enhancing both accuracy and practical utility makes LLMs more reliable in software development, saving time and reducing error potential.}
\vspace{-0.05in}
\section{Conclusion and Future Work}
This paper evaluates Code LLMs' ability to understand and generate code following design patterns. Experiments on classification, line completion, and function generation reveal that while design pattern knowledge often improves performance, some models rely on internalized patterns effectively. The study highlights challenges with complex patterns like Builder and nuances in the relationship between Code Similarity and Edit Similarity, emphasizing the need for improved design pattern integration and practical utility to enhance LLM reliability and reduce developer effort. We plan to expand the dataset to cover more object-oriented languages and incorporate more high-quality repositories. Additionally, we will observe evaluation results from more dimensions to ensure the reliability and practical guidance of our analyses.

\end{document}